\documentclass[aps,pra,showpacs, preprint]{revtex4}

\begin{document}

\title{The Bogoliubov inequality and the nature of Bose-Einstein condensates for interacting atoms in spatial dimensions $D \leq 2$ }

\author{Moorad Alexanian}
\email[]{alexanian@uncw.edu}

\affiliation{Department of Physics and Physical Oceanography\\
University of North Carolina Wilmington\\ Wilmington, NC
28403-5606\\}

\date{\today}

\begin{abstract}
We consider the restriction placed by the Bogoliubov inequality on the nature of the Bose-Einstein condensates (BECs) for interacting atoms in a spatial dimension $D\leq 2$ and in the presence of an external arbitrary potential, which may be a confining ``box," a periodic, or a disordered potential. The atom-atom interaction gives rise to a (gauge invariance) symmetry-breaking term that places further restrictions on BECs in the form of a consistency proviso.  The necessary condition for the existence of a BEC in $D\leq 2$ in all cases is macroscopic occupation of many single-particle momenta states with the origin a limit point (or accumulation point) of condensates. It is shown that the nature of BECs for noninteracting atoms in a disordered potential is precisely the same as that of BECs for interacting atoms in the absence of an external potential.
\end{abstract}


\maketitle {}

\section{Introduction} Anderson disorder-induced localization (AL) describes the sudden transition of electron mobility from that of a conductor to that of an insulator owing to the disorder in the depths of the potential wells in an otherwise periodic potential \cite{PWA58}. The noninteracting electrons are in single-particles states and it may be that the introduction of the interaction between electrons may lead to delocalization. This has prompted the study of AL of ultracold atoms in $1D$ disordered optical potentials. The ultracold atoms are considered to be noninteracting and in a BEC \cite{DC06, JB08, LF07, GR08}. This bring to the fore the fundamental question of the existence or absence of (gauge invariance) symmetry breaking terms in the Hamiltonian of interacting Bose gases in spatial dimensionality $D\leq 2$ that give rise to macroscopic occupation at finite temperatures of many single-particle momentum states.

Existing proofs of the absence of a BEC for $D\leq 2$  are based on Bogoliubov's inequality \cite{NNB61}. The proofs either (1) introduce a symmetry-breaking field into the Hamiltonian in order to get a non-vanishing order parameter at finite volume \cite{PCH67, GVC69, JFF70, AS05, ACO01}, or (2) work directly  with the static order-order correlation function (with or without a symmetry-breaking field) \cite{DJ71}. However, none of these proofs is completely rigorous \cite{MB76} since they do not take into account the infinite-dimensional nature of the Hilbert space of a Bose system in a finite volume and the consequent unbounded nature of the operators that appear in the Bogoliubov inequality. Thus, the validity of the Bogoliubov inequality itself is not established beforehand, which includes the proper handling of the thermodynamic limit \cite{MB76}. Insofar as the results of the proofs are concerned either (1) a BEC in the single-particle state with momentum \textbf{p}, usually $\textbf{p}=\textbf{0}$, is proven to be absent, or (2) a BEC in any single-particle state is shown to be absent. The former result has been established \cite{MB76} completely rigorously; while in deriving the latter results \cite{PCH67, GVC69, JFF70, AS05}, the Bogoliubov inequality is used without first proving it specifically for the infinite-dimensional case of the Bose gas.

\section{Interacting Bose gas} Consider the Hamiltonian for an interacting Bose gas
\begin{equation}
\hat{H}= \int d\textbf{r}\hat{\psi}^{\dag}(\textbf{r}) (\frac{-\hbar^2 }{2m} \nabla^2) \hat{\psi}(\textbf{r}) +  \int d\textbf{r}\hat{\psi}^{\dag}(\textbf{r}) V_{ext}(\textbf{r}) \hat{\psi}(\textbf{r}) + \int d\textbf{r} d\textbf{r}^\prime \hat{\psi}^{\dag}(\textbf{r})\hat{\psi}^{\dag}(\textbf{r}^\prime) V(\textbf{r} -\textbf{r}^\prime) \hat{\psi}(\textbf{r}^\prime)\hat{\psi}(\textbf{r}),
\end{equation}
where $V_{ext}(\textbf{r})$ is the external potential, $V(\textbf{r} -\textbf{r}^\prime)$ is the two-particle, local interaction potential, and $\hat{\psi}(\textbf{r})$ and $\hat{\psi}^{\dag}(\textbf{r})$ are bosonic field operators that destroy or create a particle at spatial position $\textbf{r}$. For the case of periodic potentials, we do not consider interactions between the bosons and the crystal that result in the creation of phonons. Macroscopic occupation in the single-particle state  $\psi(\textbf{r})$ result in the non-vanishing \cite{NNB60} of the quasi-average $\psi(\textbf{r}) = <\hat{\psi}(\textbf{r})>$ and so the boson field operator
\begin{equation}
\hat{\psi}(\textbf{r}) =\psi(\textbf{r}) + \hat{\varphi}(\textbf{r}),
\end{equation}
with
\begin{equation}
\psi(\textbf{r}) = \sqrt{\frac{N_{0}}{V(D)}}\sum_{\textbf{k}^\prime} \xi_{\textbf{k}^\prime} e^{i \textbf{k}^\prime \cdot\textbf{r}} \equiv\sqrt{\frac{N_{0}}{V(D)}} f(\textbf{r}) ,
\end{equation}
and
\begin{equation}
\sum_{\textbf{k}^\prime}|\xi_{\textbf{k}^\prime}|^2 =1,
\end{equation}
where $N_{0}$ is the number of atoms in the condensate and $V(D)$ is the D-dimensional "volume" and $<\hat{\varphi}(\textbf{r})> =0$.  The operator $\hat{\varphi}(\textbf{r})$ has no single-particle states that are in the condensate and so $\int d\textbf{r} \hat{\varphi}^\dag(\textbf{r}) \psi(\textbf{r}) = 0$. The separation of $\hat{\psi}(\textbf{r})$ into two parts gives rise to the following (gauge invariance) symmetry breaking term in the Hamiltonian (1)
\[
\hat{H}_{symm} = \int d\textbf{r} \hat{\varphi}^\dag(\textbf{r}) \psi(\textbf{r})  \int d\textbf{r}^\prime
 [V(\textbf{r} - \textbf{r}^\prime) + V(\textbf{r}^\prime - \textbf{r})] |\psi(\textbf{r}^\prime)|^2   + h. c.
 \]
\begin{equation}
\equiv \int d\textbf{r} \hat{\varphi}^\dag(\textbf{r}) \chi(\textbf{r}) + h. c.
\end{equation}

The presence of this nonzero $\hat{H}_{symm}$ in the Hamiltonian gives rise to further macroscopic occupation in states other than the original state given by $\psi(\textbf{r})$ and so the condensate wavefunction $\psi(\textbf{r})$ gets modified by augmenting the single-particles states where macroscopic occupation occurs. In such a case, macroscopic occupation in the state $b$ would give rise to macroscopic occupation in the states $a$, such that $a\neq b$, whenever  the matrix element $<ab|\hat{V}|bb>$  of the potential $\hat{V}$, which is the last term in Eq. (1), does not vanish. For noninteracting particles in an external potential, the Hamiltonian is diagonal in the representation of the eigenstates of the Hamiltonian and so macroscopic occupation in any given eigenstate of the Hamiltonian does not give rise to further macroscopic occupation in any other energy eigenstate. However, the effect of interparticle interactions can generate macroscopic occupation in other energy eigenstates. For instance, for a harmonic trap, macroscopic occupation of the lowest energy state, the ground state, does not give rise to macroscopic occupation of any of the higher energy eigenstates. However, interparticle interactions may allow the generation of macroscopic occupation in other energy eigenstate thus modifying the original BEC. Note, however, that macroscopic occupation only in the single-particle state with momentum $\textbf{p}$ does not give rise to macroscopic occupation in any other momentum state since the matrix element in the momentum representation $<\textbf{q}\textbf{p}| \hat{V}|\textbf{p}\textbf{p}>$ vanishes by momentum conservation unless $\textbf{q}= \textbf{p}$.

This consistency proviso requires that the correct condensate wavefunction $\psi(\textbf{r})$ corresponds to that which gives rise to no symmetry breaking term in the Hamiltonian. That is to say, $\hat{H}_{symm}$ vanishes for the correct condensate wavefunction $\psi(\textbf{r})$.  For instance, macroscopic occupation in the single-particle states with momenta $\textbf{0}, \pm\textbf{k}_{0}$ gives rise \cite{MA78}, with the aid of the symmetry breaking term $\hat{H}_{symm}$ and owing to linear momentum conservation, to macroscopic occupation in the single-particle momenta states $ \pm 2 \textbf{k}_{0}$; therefore, $\hat{\varphi}^\dag(\textbf{r})$ is orthogonal to both $\psi(\textbf{r})$ and $\chi(\textbf{r})$ and thus one has the possibility of macroscopic occupation in all the momentum states $\textbf{0}, \pm\textbf{k}_{0}, \pm 2 \textbf{k}_{0}, \pm 3 \textbf{k}_{0}, \cdots$. In general, one can have also a finite or an infinite sum \cite{MA78} over the momentum variable $\textbf{k}_{0}$ and so the momentum conserving two-particle interaction in the Hamiltonian (1) allows for condensates in the single-particle momentum states $\textbf{k} + \sum_{i} m_{i}\textbf{k}_{i}$, where $m_{i} = 0, \pm1, \pm 2, \cdots$ \cite{MA81}. The momenta $\{ \textbf{k}_{i}\}$ represent a set of vectors that are, in general, incommensurate thus giving rise to nonperiodic or aperiodic condensates. However, if one or more of the momenta $\textbf{k}_{i}$ approaches zero, then the sequence of condensates over the momenta sequence $\{\textbf{k}^\prime\}$ with nonzero $\xi_{\textbf{k}^\prime}$ has $\textbf{k}^\prime = 0$ as a limit point (or accumulation point) as $\textbf{k}_{i} \rightarrow 0$ thus removing the $1/k^2$--singularity \cite{MA80} that is necessary in all proofs of the absence of a BEC for $D \leq 2$. For a system confined to a box of length $L$ by an external potential, momenta is quantized and so $\textbf{k} = \frac{2\pi}{L} n_{x}\hat{\textbf{x}} + \frac{2\pi}{L} n_{y}\hat{\textbf{y}} + \frac{2\pi}{L} n_{z}\hat{\textbf{z}}$.

Note, however, that recent papers still suppose that interacting Bose gases in $D \leq 2$ and at $T>0$ possess no BECs \cite{KHD07,CCB09, WZ08}. The phase transition is described \cite{KHD07} by the Berezinskii-Kosterlitz-Thouless theory \cite{VLB72, KT73, LSH05}, which does not involve any spontaneous symmetry breaking but, instead, is associated with a topological order embodied in the pairing of vortices with opposite circulations. It is interesting that both the singly quantized vortex in a doubly connected system and the condensate that gives rise to two vortices of equal and opposite circulations are both described by condensates with a macroscopic occupation of infinitely many single-particle states with $k=0$ a limit-point \cite{MA80}. Similarly, limits on Bose-Einstein condensation in confined solid $^{4}\textup{He}$, where large superfluid fractions have been reported, is based \cite{DAK09} on supposing macroscopic occupation in only the single-particle momentum state $\textbf{k} =0$, which is not the most general, possible BEC. In addition, it is supposed that for $2D$, superfluidity is not a consequence of a BEC \cite{VLB72, KT73} but is associated with the onset of algebraically decaying off-diagonal long-range order (ODLRO) \cite{Y62}. However, the existence of ODLRO can be a consequence of macroscopic occupation in many single-particle momenta states, which is equivalent to the existence of a BEC. In what follows, we consider macroscopic occupation for $D\leq 2$ in many-momenta states with $\textbf{k}=0$ a point of accumulation.

\subsection{ Example of BEC in $1D$} Consider the following one-dimensional example for a BEC with macroscopic occupation in the single-particle momentum states $k=k_{0}n$ with $n=0, \pm1, \pm2, \cdots$ for a Bose system in the presence of a nonperiodic potential, where $k_{0} = 2\pi/L$. The BEC (3) becomes
\begin{equation}
\psi(x) = \sqrt{ \frac {N_{0}}{2\pi}}\hspace{0.05in} \Big( \sum_{n=-\infty}^{+\infty} \frac{k_{0}}{(k_{0}^2 n^2 + \kappa^2)^{2 \nu +1}  } \Big) ^{-1/2} \sum_{ n = -\infty}^{+\infty}\frac{k_{0} \textup{cos}(k_{0}n x)}{( k_{0}^2 n^2 + \kappa^2)^{\nu + 1/2}}, \hspace{0.1in} (0 \leq x \leq  2 \pi/k_{0}),
\end{equation}
where $\nu > 0$ so that $\psi(0)$ is finite.  The BEC is normalized as follows  $\int_{0}^{2\pi/k_{0}} \textup{d}x |\psi (x)|^2  = N_{0}$. Expansion (6) represents a Fourier series and so $\psi(x + 2\pi/k_{0}) = \psi(x)$, which is a dynamical consequence since macroscopic occupation in the single-particle momentum states $k= 0, \pm k_{0}$ implies macroscopic occupation in the states with momenta $k = n k_{0}$ with $ n= 0, \pm 1, \pm 2, \cdots$ owing to the symmetry-breaking term (5). As a function of the complex variable $k_{0}$, the Fourier series in (6) possesses simple or higher order poles at $k_{0} = \pm i\kappa/n$, $n=1, 2, 3, \cdots$,  for $\nu+1/2$ = positive integer or branch point singularities at $k_{0} = \pm i\kappa/n$, $n=1, 2, 3, \cdots$,  for $\nu+1/2 \neq$ positive integer. Therefore, $k_{0} = 0$ is an accumulation point of poles or branch point singularities depending on the value of $\nu$.

The sum (6) can be approximated with great accuracy by an integral when $k_{0}$ is arbitrarily small. This would represent the passage of the Fourier series for the periodic function to a Riemann integral for a nonperiodic function. Accordingly, in the limit $k_{0} \rightarrow 0$, $k=0$ becomes a point of accumulation of condensates, the sum (6) may be converted into an integral, and so
\[
\psi(x)= \sqrt{ \frac {N_{0}}{2\pi}}\hspace{0.05in} \Big( \int_{-\infty}^{+\infty} \textup{d}k \frac{1}{(k^2  + \kappa^2)^{2 \nu +1}  } \Big) ^{-1/2} \int_{ -\infty}^{+\infty} \textup{d}k\frac{ \textup{cos}(k x)}{( k^2  + \kappa^2)^{\nu + 1/2}}
\]
\begin{equation}
=\frac{1}{2^{\nu -1/2}  \pi^{1/4}} \sqrt{  \frac{\kappa \Gamma(2\nu + 1) N_{0}}{\Gamma(2\nu +1/2) \Gamma^{2}(\nu + 1/2)}} \hspace{0.1in} \kappa^{\nu} |x|^{\nu} K_{\nu}(\kappa |x|),  \hspace{0.1in} (-\infty < x < \infty),
\end{equation}
where $\Gamma (z)$ is the gamma function and $K_{\nu}(z)$ is the modified Bessel function of the second kind. The integral (7) converges for $|x| > 0$ provided $\nu>-1/2$; however, if the condensate wavefunction is required to be  bounded at $x=0$, then (7) converges for $|x|\geq 0$ provided $\nu > 0$.

Now  $\textup{d} [z^{\nu} K_{\nu}(z)]/\textup{d} z =- z^{\nu } K_{\nu -1}(z)$, $K_{\nu}(z) \rightarrow 2^{\nu -1} \Gamma(\nu)/z^{\nu}$ as $z\rightarrow 0$ for $\Re \nu>0$, and  $K_{-\nu}(z) = K_{\nu}(z)$. Therefore, $z^\nu K_{\nu}(z) = 2^{\nu -1} \Gamma(\nu) - 2^{\nu -3} \Gamma(\nu -1) z^2 + \cdots$ as $z \rightarrow 0$ for $\nu >1$ and so we have a local maximum at the middle of the BEC given by (7). Next one has that  $z^\nu K_{\nu}(z) = 2^{\nu -1} \Gamma(\nu) - \Gamma(1- \nu)z^{2\nu}/(2^{\nu +1} \nu)   + \cdots$ as $z\rightarrow 0$ for $0<\nu<1$ and so for $0<\nu<1/2$ the BEC (7) has a cusp singularity at $x=0$ with the derivative tending toward $\infty$  or $-\infty$  as one approaches the cusp. For $\nu =1/2$, the BEC (7) is continuous at $x=0$ but the derivative is discontinuous there with  slope of $ \sqrt{\pi/2}$ or $- \sqrt{\pi/2}$ as one approaches $x=0$. For $ 1/2 <\nu <1$, the BEC (7) attains its largest value of the center of the localized BEC but it is not a local maximum. The extremal cases of $\nu =0, 1$ are as follows. For $\nu =0$, $\psi(x) \propto K_{0}(\kappa |x|) = -\ln |x| +\cdots$  thus there is a cusp singularity with slope $\infty$ or $-\infty$ as one traverses $x=0$. Notice that BEC (7) is not bounded at the center of the localized condensate for $\nu =0$; nonetheless, the integral of the  BEC density, which gives the total number of particles in the condensate, is finite. Finally, for $\nu=1$, $ z K_{1}(z) = 1 + (z^2/2) \ln z + \cdots$ as $z\rightarrow 0$ and so $\psi(x)$  attains its largest value at $z=0$ but it is not a local maximum.

The standard deviations $\Delta x$ and $\Delta p$ follow directly from (7)
\begin{equation}
\Delta x = \kappa \sqrt{\frac { (\nu+1/4)(\nu + 1/2)}{ \nu +1}}, \hspace{.5in} \Delta p = \frac{\hbar }{2 \kappa} \sqrt{\frac { 1}{ \nu -1/4}},
\end{equation}
and so
\begin{equation}
(\Delta x)(\Delta p) = \frac{\hbar}{2}  \sqrt{  \frac{(\nu + 1/4)(\nu + 1/2)}{(\nu +1)(\nu -1/4)}} \geq \frac{1}{2} \hbar,
\end{equation}
for $\nu > 1/4$.

The sum in (6) can be carried out explicitly for $\nu=1/2$ and one obtains
\begin{equation}
\psi (x) = A \hspace{0.05in} \frac{ \pi }{\kappa}\sqrt{ \frac {N_{0}}{2\pi}}  \hspace{0.07in} \Big( \frac {e^{\pi\kappa/k_{0}} e^{-\kappa x} + e^{-\pi\kappa/k_{0}} e^{\kappa x}}{e^{\kappa \pi/k_{0}} - e^{-\kappa \pi/k_{0} }}\Big), \hspace{0.5in}  (0 \leq x \leq 2\pi/k_{0} ),
\end{equation}
where $\phi(x + 2\pi/k_{0}) = \phi (x)$ and the normalization constant $A$ is
\begin{equation}
A= \Big(\frac{\pi}{2\kappa^3} \coth(\pi \frac{\kappa }{k_{0}})+   \frac{\pi^2}{2 k_{0} \kappa^2} \textup{csch}^{2}(\pi \frac{\kappa }{k_{0}})  \Big)^{-1/2}.
\end{equation}
Note that for $k_{0} \rightarrow 0$, one has that
\begin{equation}
\psi (x) = \sqrt{\kappa N_{0}} \hspace{0.05in} e^{-\kappa|x|},    \hspace{.3in}        (-\infty < x <\infty),
\end{equation}
which agrees with Eq. (7) since $K_{1/2}(z) = \sqrt{\pi/2 z} \hspace{0.06in} e^{-z}$. Result (12) corresponds to a localized BEC that decays exponentially and has a discontinuous derivative at the center.

For $\nu>0$, $K_{\nu}(z) \rightarrow 2^{\nu -1}\Gamma (\nu)/z^\nu$ as $z\rightarrow 0$ and so the condensate wavefunction  $\psi(x)$ given by (7) is finite at $x=0$. In addition, $K_{\nu}(z) \rightarrow \sqrt{\pi/2z} \hspace{0.07in} e^{-z}$ as $z\rightarrow \infty$ for $\nu > -1/2$ and so the condensate wavefunction (7) is similarly localized.
Note that prior to taking the limit $k_{0} \rightarrow 0$, which approximates the sum (6) with the integral (7), the condensate wavefunction is periodic, viz.,  $\psi ( x + 2\pi/k_{0}) = \psi(x)$. However, in the limit $k_{0} \rightarrow 0$, one has an even, nonperiodic, localized condensate wavefunction.

Actually, one can interchange the roles of $x$ and $k$ in Eq. (7) and so one has, instead, a condensate wavefunction $\psi (x) \propto (x^2 + a^2)^{-\nu -1/2}$ with a corresponding momentum distribution $\varphi (k) \propto |k|^\nu K_{\nu}(a|k|)$, where $a$ is a length scale. It is interesting that in experiments with ultracold atoms (Ref. 28), BECs have been found that suggest a power-law decrease in the wings of the atomic density with an exponent close to the value 2, viz., $|\psi(x)|^{2} \propto x^{-2}$ as $|x|\rightarrow \infty$. This would correspond to $\nu =0$ and so $|\psi(x)|^2 \propto (x^2 + a^2)^{-1}$ for $(-\infty < x < \infty)$. The momentum distribution associated with this BEC density is given by the modified Bessel function of order zero $\varphi (k) \propto K_{0}(a|k|)$ for $(-\infty < k < \infty)$, where the singular point $k=0$ is a  logarithmic branch point since $K_{0}(z) \propto -\textup{ln} z$ as $z \rightarrow 0$. The standard deviation $\Delta p = \hbar/(2 a \sqrt{2})$ is finite even though $\varphi(k)$ diverges logarithmically as $k\rightarrow 0$; however, $\Delta x $ is infinite even though $|\psi (x)|^2$ is bounded everywhere. Note that the behavior of $\psi(x)$ for $x \gg a $ is determined by the behavior of its Fourier transform $\varphi (k)$ for $k a\ll 1 $.

It is interesting to consider a finite rather than an infinite sum in Eq. (6),
\begin{equation}
\chi(x) = \sqrt{ \frac {N_{0}}{2\pi}}\hspace{0.05in} \Big( \sum_{n=-N/2}^{ N/2} \frac{k_{0}}{(k_{0}^2 n^2 + \kappa^2)^{2 \nu +1}  } \Big) ^{-1/2} \sum_{ n = - N/2}^{N/2}\frac{k_{0} \textup{cos}(k_{0}n x)}{( k_{0}^2 n^2 + \kappa^2)^{\nu + 1/2}}, \hspace{0.1in} (0 \leq x \leq  2 \pi/k_{0}),
\end{equation}
and thus consider  the combined limit $k_{0} \rightarrow 0$ and $N\rightarrow \infty$ in order to approximate the sum by an integral. If $k_{0}N \rightarrow \infty$ as $k_{0} \rightarrow 0$ and $N\rightarrow \infty$, then one obtains the previous result given by Eq. (7). However, if $k_{0}N/2 \rightarrow K <\infty$ as $k_{0} \rightarrow 0$ and $N\rightarrow \infty$, then
\begin{equation}
\chi(x)= \sqrt{ \frac {N_{0}}{2\pi}}\hspace{0.05in} \Big( \int_{- K}^{ K} \textup{d}k \frac{1} {(k^2 + \kappa^2)^{2 \nu +1}} \Big)^{-1/2}   \int_{-K}^{K} \textup{d}k \frac{ \textup{cos}(xk)}  { (k^2 + \kappa^2 )^{\nu + 1/2}}, \hspace{0.3in} (-\infty < x <\infty).
\end{equation}
For one-dimensional crystals, $K \leq\pi/a $, where $a$ is the length of a unit cell. The condensate wavefunction  $\chi(x)$ is normalized by $\int_{-\infty}^{+\infty} \textup{d}x |\chi(x)|^2 = N_{0}$. Note that for $ K |x| \gg 1$, the second integral in (14) is dominated by the small values of $k$, viz., $k \ll K$, and so the range of the integral in (14) can be extended to $\pm \infty$ when  $ K |x| \gg 1$ and one has that $\chi(x)\propto |x|^{\nu} K_{\nu}(\kappa |x|)$ for $ K |x| \gg 1$, which becomes $\chi(x) \propto  |x|^{\nu -1/2} \hspace{0.05in} e^{-\kappa |x|}$ for $\kappa |x| \gg 1$.

\subsection{BEC in the harmonic trap}

If the external potential is a one-dimensional harmonic oscillator, then the field operator $\hat{\psi}(x)$ is expanded in terms of the energy eigenstates $\psi_{n}(x)$ of the harmonic oscillator. If one has macroscopic occupation in the ground state $\psi_{0}(x)$, then $\hat{\psi}(x) = \psi_{0}(x) + \hat{\varphi}(x)$, where the operator $\hat{\varphi}(x)$ is orthogonal to $\psi_{0}(x)$ and so $\hat{\varphi}(x)$ has nonzero expansion coefficients for only the creation operators of the higher energy harmonic oscillator eigenstates. It should be noted that if one considers atom-atom interactions, then it may be that states other than the ground state may be also macroscopically occupied. The latter occurs if the matrix element $< l 0|\hat{V}|0 0>$  of the atom-atom interaction potential $\hat{V}$ does not vanish for $l \neq 0$ in which case there would be additional macroscopic occupations in the higher energy harmonic oscillator eigenstates $\psi_{l}(x)$.

Consider the following illustration of a condensate in the ground state of a one-dimensional harmonic trap
\begin{equation}
\psi_{0}(x) = \sqrt{\frac{N_{0}}{2\pi}} \Big(\sum_{n = -\infty}^{\infty}  k_{0} \hspace{0.03in} e^{-2\beta k_{0}^2 n^2}\Big)^{-1/2}  \sum_{n = -\infty}^{\infty}  k_{0} \hspace{0.03in} e^{-\beta k_{0}^2 n^2} \textup{cos} (k_{0} nx),   \hspace{0.5in}  -\pi/k_{0} \leq x  \leq \pi/k_{0}.
\end{equation}
The condensate (15) has macroscopic occupation in the single-particle momentum states $k_{0} n$, with $n= 0, \pm1, \pm 2, \cdots$, $\psi(x)$ is periodic $\psi(x + 2\pi/k_{0}) = \psi(x)$, and is normalized as follows $\int_{-\pi/k_{0}}^{\pi/k_{0}}  \textup{d}x |\psi(x)|^2 = N_{0}$. Each Fourier coefficient of the series (15) possesses an essential singularity at $k_{0} =\infty$ and in the limit $k_{0} \rightarrow 0$, may be approximated arbitrarily well by a Riemann integral and so
\[
\psi_{0}(x) = \sqrt{\frac{N_{0}}{2\pi}} \Big(\int_{ -\infty}^{\infty}  \textup {d}k \hspace{.04in} e^{-2\beta k^2}\Big)^{-1/2}  \int_{ -\infty}^{\infty}  \textup{d}k   \hspace{0.1in} e^{-\beta k^2} \textup{cos} (k x)
\]
\begin{equation}
= \sqrt{N_{0}} \Big(\frac{1}{2\pi\beta}\Big)^{1/4}   e^{-x^2/4\beta},  \hspace{0.5in}  -\infty < x < \infty.
\end{equation}
Therefore, the condensate associated with the lowest energy state in the harmonic trap is represented by the macroscopic occupation of single-particle momentum states given by $k=nk_{0}$, where $n=0, \pm 1, \pm 2, \cdots$, with $k=0$ a point of accumulation as $k_{0}\rightarrow 0$.

Note that as a function of the complex variable $k_{0}$, the BEC given by the series (15) converges for $\Re k_{0}^2 = (\Re k_{0})^2 - (\Im k_{0})^2 > 0$ and diverges for  $\Re k_{0}^2 \leq 0$. The regions of convergence and divergence resemble a Minkowski spacetime diagram. The series (15) converges in the spacelike regions and diverges in the timelike regions and the light cones. Along the real axis, the series (15) converges except the singularity that it encounters at the origin $k_{0} =0$.

\section{ Bogoliubov inequality} The absence or presence of a BEC in spatial dimensions $D \leq 2$ is based on Bogoliubov's inequality
\begin{equation}
\frac{1}{2}\langle\{\hat{A},\hat{A}^{\dag}\}\rangle \geq k_{B}T |\langle[\hat{C},\hat{A]}\rangle|^2 / \langle[[\hat{C}, \hat{H}], \hat{C}^{\dag}]\rangle,
\end{equation}
where $\hat{H}$ is the Hamiltonian (1) of the system with arbitrary local interparticle and external potentials, the brackets denote thermal averages, and the operators $\hat{A}$ and $\hat{C}$ are arbitrary provided all averages exist.

Consider the following operators \cite{MA80},
\begin{equation}
\hat{C} = \int d\textbf{r} e^{i\textbf{k}\cdot \textbf{r}} \hat{\psi}^{\dag}(\textbf{r})\hat{\psi}(\textbf{r})
\end{equation}
and
\begin{equation}
\hat{A} = \int d \textbf{r} \int d \textbf{r}^{\prime} e^{-i\textbf{k}\cdot \textbf{r}} f(\textbf{r}) f^{\ast}(\textbf{r}^\prime) \hat{\psi}^{\dag}(\textbf{r})\hat{\psi}(\textbf{r}^\prime),
\end{equation}
where $\textbf{k}$ is arbitrary. Now,
\begin{equation}
\langle [A^\dag,A]\rangle = V(D)\langle[\hat{C},\hat{A}]\rangle = V^2(D)\{ N_{0} - N_{0} |A_{\textbf{k}}|^2 -\sum_{\textbf{q}}\langle \hat{a}^\dag_{\textbf{q}} \hat{a}_{\textbf{q}}\rangle  | \xi_{\textbf{q}+ \textbf{k}}|^2 \},
\end{equation}
\begin{equation}
\langle[[\hat{C}, \hat{H}], \hat{C}^{\dag}]\rangle = \frac{\hbar^2 k^2}{m} N,
\end{equation}
with
\begin{equation}
A_{\textbf{k}} = \sum_{\textbf{k}^\prime} \xi_{\textbf{k}^\prime} \xi^{*}_{\textbf{k}^\prime + \textbf{k}}
= \frac{1}{N_{0}} \int \textup{d} \textbf{r} |\psi(\textbf{r})|^2 e^{i \textbf{k}\cdot \textbf{r}}
\end{equation}
with the aid of Eq. (3), where $N$ is the total number of particles. Now, $|A_{\textbf{k}}| \leq 1$  by the Cauchy-Schwarz inequality  where the equality holds for $\textbf{k} = 0$, that is, $A_{0}=1$ with the aid of Eq. (4). The vector $\textbf{q} \notin \{\textbf{k}^\prime\}$, where $\{\textbf{k}^\prime\}$ is the set of condensate vectors for which $\xi_{\textbf{k}^\prime} \neq 0$.  For a BEC at rest, $|\xi_{\textbf{k}}|^2 = |\xi_{- \textbf{k}}|^2$. Note that $\xi_{\textbf{q} + \textbf{k}}\neq 0$ for $(\textbf{q} + \textbf{k}) \in \{\textbf{k}^\prime\}$ and $\xi_{\textbf{q} + \textbf{k}}= 0$ for $\textbf{q}\notin \{\textbf{k}^\prime\}$ and $ \textbf{k} \in \{\textbf{k}^\prime\}$.

We sum the Bogoliubov inequality (17) over the single-particle momentum states in the set $\{\textbf{k}^\prime\}$ constituting the condensate, which includes an arbitrary neighborhood of the point of accumulation of the condensate at $\textbf{k}^\prime = 0$ that corresponds to a condensate at rest. We want to find an upper bound of the anticommutator   $\langle\{\hat{A},\hat{A}^{\dag}\}\rangle = 2\langle \hat{A}\hat{A}^{\dag}\rangle + \langle [\hat{A}^{\dag},\hat{A}]\rangle$. We extend the sum over the first term $\langle \hat{A}\hat{A}^{\dag}\rangle$ over all values of $\textbf{k}$ thus obtaining a larger upper bound
\begin{equation}
M N_{0}N V^2(D) \geq \sum_{\textbf{k}}\langle \hat{A}\hat{A}^{\dag}\rangle,
\end{equation}
where the condensate wavefunction $\psi(\textbf{r})$ is orthogonal to the operator $\hat{\varphi}^\dag(\textbf{r})$ and we assume that the condensate density is bounded from above by $|f(\textbf{r})|^2 \leq M$ with $1 \leq M <\infty$ since $\int d \textbf{r}|f(\textbf{r})|^2 =V(D)$. In (23) use has been made of the completeness relation for the momentum eigenstates and a negative term resulting from a single commutation has been dropped. Note that we are considering a condensate where all the single-particle states with momentum $\textbf{k}^\prime$ are occupied macroscopically with $\textbf{k}^\prime = \textbf{0}$ a point of accumulation. In addition, we are supposing that the number of particles in the ``volume" $V(D)$ is fixed, that is, we are employing a canonical ensemble and so  $\int d\textbf{r}\hat{\psi}^{\dag}(\textbf{r})\hat{\psi}(\textbf{r}) = \sum_{\textbf{k}} \hat{a}^\dag_{\textbf{k}} \hat{a}_{\textbf{k}} =  \hat{N}$ is actually the c-number $N$.

Consider next the sum over $\textbf{k}^\prime$ of the commutator $\langle[\hat{A}^\dag,\hat{A]}\rangle$,
\begin{equation}
\sum _{\textbf{k}^\prime} \langle[\hat{A}^\dag,\hat{A]}\rangle = N_{0} V^2(D)  \sum _{\textbf{k}^\prime} [ 1 - |A_{\textbf{k}^\prime}|^2],
\end{equation}
with the aid of (20) and where $\xi_{\textbf{q} + \textbf{k}} =0$ for $\textbf{q}\notin \{\textbf{k}^\prime\}$ and $\textbf{k}\in \{\textbf{k}^\prime\}$. This sum over the commutator is bounded from above provided the sum is restricted to values of $\textbf{k}^\prime $ that have a finite, upper bound. Now the right-hand side (RHS) of inequality (17) becomes
\begin{equation}
k_{B}T |\langle[\hat{C},\hat{A]}\rangle|^2 / \langle[[\hat{C}, \hat{H}], \hat{C}^{\dag}]\rangle = \frac{m k_{B}T}{\hbar^2} \frac{ V^2(D) N_{0}^2}{N} \sum_{\textbf{k}^\prime}\Bigl ( \frac{1 - |A_{\textbf{k}^\prime}|^2}{k^\prime} \Bigr )^2
\end{equation}
with the aid of Eqs. (20) and (21). Note that the RHS is bounded in the upper limit of the sum; however, it is the lower limit as $\textbf{k}^\prime \rightarrow 0$ for $D \leq 2$ where the sum may diverge which would result in no BECs, viz., $N_{0}=0$ for $T>0$. Combining Eqs. (23)--(25), we have for the Bogoliubov inequality,
\begin{equation}
M N + \frac{1}{2} \sum_{\textbf{k}^\prime}( 1 - |A_{\textbf{k}^\prime}|^2)  \geq \frac{m k_{B}T }{\hbar^2} \frac{ N_{0}}{N} \sum_{\textbf{k}^\prime}\Bigl( \frac{1 - |A_{\textbf{k}^\prime}|^2}{k^\prime} \Bigr) ^2.
\end{equation}
 Therefore, the existence of a BEC for $T>0$ requires the convergence of the sum on the RHS of (26) over the macroscopically occupied single-particle momentum states $\textbf{k}^\prime$ of the condensate. Note that the sums in (26) over the condensate momenta can be approximated by integrals according to $\sum_{\textbf{k}^\prime} \rightarrow V(D) \int \textup{d} \textbf{k}^\prime$ and so
 \begin{equation}
M \frac{ N}{V(D)} + \frac{1}{2} \int \textup{d} \textbf{k}^{\prime}( 1 - |A_{\textbf{k}^\prime}|^2)  \geq \frac{m k_{B}T }{\hbar^2} \frac{ N_{0}}{N} \int \textup{d} \textbf{k}^{\prime} \Bigl( \frac{1 - |A_{\textbf{k}^\prime}|^2}{k^\prime} \Bigr) ^2.
 \end{equation}
 The integral on the RHS has no infrared divergence since by (22), $(1 -|A_{\textbf{k}^\prime}|^2)$ vanishes quadratically as $\textbf{k}^\prime \rightarrow  0$ and so the $1/k^2$--singularity is removed thus allowing the existence of a BEC for $D\leq 2$.

\subsection{Removal of $1/k^2$--singularity in $1D$} To illustrate the removal of the $1/k^2$--singularity for the existence of a BEC for $D\leq 2$, consider the  $1D$ example given by Eq. (6) where
\begin{equation}
\xi_{k_{0}n} =  \frac{B }{(k_{0}^2 n^2 + \kappa^2)^{\nu + 1/2}} \hspace{0.5in}  (n= 0, \pm 1, \pm 2, \cdots),
\end{equation}
with the normalization constant $B$ given by
\begin{equation}
B = (\sum_{n = -\infty}^{+ \infty} \frac{1}{(k_{0}^2 n^2 + \kappa^2)^{2 \nu + 1}})^{-1/2}
\end{equation}
with the aid of (4) and so by (22)
\begin{equation}
A_{k_{0}l} = A_{\textup{-}k_{0}l} =  B^2 \sum_{n = -\infty}^{+ \infty} \frac{1}{(k_{0}^2 n^2 + \kappa^2)^{\nu + 1/2}} \hspace{0.04in} \frac{1}{(k_{0}^2 (n - l)^2 + \kappa^2)^{ \nu + 1/2}}.
\end{equation}
Now
\[
\frac{\Gamma( \nu + 1/2)}{[k_{0}^2 (n - l)^2 + \kappa^2]^{ \nu + 1/2}} = \int_{0}^{\infty} \textup{d}x \hspace{0.04in} x^{\nu - 1/2} \hspace{0.04in} e^{-[k_{0}^2 (n - l)^2 + \kappa^2] x}
\]
\begin{equation}
= \sum_{j=0}^{\infty} (-1)^j \hspace{0.04in}   \frac{\Gamma(j+\nu + 1/2) \hspace{0.04in}k_{0}^{2j} (l^2 - 2n l)^{2j}}{ j! (k_{0}^2 n^2 + \kappa^2)^{j+\nu+ 1/2}}
\end{equation}
for $\nu>-1/2$, which gives the following series expansion for (30) in the neighborhood of $l=0$,
\[
A_{k_{0}l} = 1 - B^2  \frac{\Gamma(\nu + 3/2)}{\Gamma(\nu + 1/2)}  \sum_{n = -\infty}^{+\infty} \frac{k_{0}^2 l^2}{(k_{0}^2 n^2 +\kappa^2)^{2\nu +2}} +
 \]
\begin{equation}
 + 2 B^2  \frac{\Gamma(\nu + 5/2)}{\Gamma(\nu + 1/2)}  \sum_{n = -\infty}^{+\infty} \frac{k_{0}^4  l^2 n^2}{(k_{0}^2 n^2 +\kappa^2)^{2\nu +3}} + O (l^4).
\end{equation}
The sum on the RHS of (26) is
\begin{equation}
\sum_{l= -\infty}^{+\infty}   \frac{1}{k_{0}^2 l^2} (1 - A_{k_{0}l})^2  (1 + A_{k_{0}l})^2
\end{equation}
and so the singularity at $l=0$ in the summand is removed as indicated by the expansion (32) of $A_{k_{0}l}$.

Similarly for the BEC in the harmonic trap where (15) gives that
\begin{equation}
A_{k_{0}l} = \Big( \sum_{n= -\infty}^{\infty} e^{-2\beta k_{0}^2 n^2} \Big)^{-1}  \sum_{n=-\infty}^{\infty} e^{-2\beta k_{0}^2 n^2} e^{-\beta k_{0}^2 l^2} \cosh (2 n \beta l k_{0}^2),
\end{equation}
which on expanding in powers of $l$ about $l=0$ removes the $k_{0}^2 l^2$--singularity at $k_{0}l = 0$ in the series on the RHS of the Bogoliubov inequality (26).

\section{BECs in periodic and disordered potentials} The existence of superfluidity is usually associated with the existence of a BEC. The existence of superflow \cite{EK04} in solid helium $^4 \textup{He}$ has stimulated the search of a BEC in solid helium thus establishing the existence of BECs in all three states of matter--gas, liquid, and solid. It is to be noted that the proofs of the absence of a BEC in gases and liquids for $D \leq 2$ applies also for atoms in external infinite periodic potentials and thus to crystalline solids. If one supposes macroscopic occupation in the momenta states $\textbf{k}, \textbf{k} \pm\textbf{q}_{0}$, then the symmetry breaking term (5) gives rise to macroscopic occupation in the momenta states $\textbf{k} \pm n \textbf{q}_{0}$ with $n= 0, \pm 1, \pm 2, \cdots$. Accordingly, the condensate wavefunction is given by
\begin{equation}
\psi_{\textbf{k}}(\textbf{r}) = \sqrt{\frac{N_{0}}{V(D)}}\sum_{n= -\infty}^{ + \infty} \xi_{\textbf{k}+ n \textbf{q}_{0} } e^{i (\textbf{k}+ n \textbf{q}_{0})\cdot\textbf{r}} \equiv  e^{i \textbf{k}\cdot \textbf{r}} u_{\textbf{k}}(\textbf{r}),
\end{equation}
which is of the Bloch form since $u_{\textbf{k}}(\textbf{r})$ has the periodicity of the lattice, that is,  $u_{\textbf{k}}(\textbf{r}) = u_{\textbf{k}}(\textbf{r} + \textbf{t}_{m})$ since $e^{i \textbf{q}_{0}\cdot \textbf{t}_{m}} = 1$. The primitive lattice translation vector   $\textbf{t}_{m} = m_{1} \textbf{a} + m_{2} \textbf{b} + m_{3} \textbf{c}$, where $m_{i}$ can take all integer values and $\textbf{a}$, $\textbf{b}$, and $\textbf{c}$ are the edges of the unit cell, which forms a parallelepiped. Therefore, the possible values of the lattice condensate vector(s)  $\textbf{q}_{0}$ are in \textbf{q}-space or reciprocal space. Note that Bloch functions $\psi_{\textbf{k}}(\textbf{r})$ and $\psi_{\textbf{k}^\prime}(\textbf{r})$ are orthogonal for $\textbf{k}^\prime \neq \textbf{k}$ provided $|\textbf{k}| < |\textbf{q}_{0}|/2$ and $|\textbf{k}^\prime| < |\textbf{q}_{0}|/2$ and so $\int \textup{d}\textbf{r} \psi_{\textbf{k}^\prime}^*(\textbf{r}) \psi_{\textbf{k}}(\textbf{r}) = N_{0} \delta_{\textbf{k}^\prime, \textbf{k}}$.  The expectation value of the momentum in the condensate follows from (35) and so
\begin{equation}
\int \textup{d}\textbf{r} \psi^\ast_{\textbf{k}}(\textbf{r})(-i \hbar \nabla) \psi_{\textbf{k}}(\textbf{r}) = N_{0}\hbar \sum_{n=-\infty}^{\infty} (\textbf{k} +n \textbf{q}_{0})|\xi_{\textbf{k} + n \textbf{q}_{0}}|^2 = N_{0} \hbar \textbf{k}
\end{equation}
provided $|\xi_{\textbf{k} + n \textbf{q}_{0} }|^2 = |\xi_{\textbf{k} - n \textbf{q}_{0} }|^2$. Therefore, $\textbf{k}$ represents the energy dependent Bloch wave number or the quasi momentum per particle $\hbar  \textbf{k}$ of the condensate.

For the case of a $1D$ crystal of length $L = Na$, where there are $N$ primitive cells of length $a$ one has that
\begin{equation}
\psi_{k}(x) = e^{ikx}u_{k}(x),
\end{equation}
with $u_{k}(x+a)= u_{k}(x)$, where $k = 2\pi m/Na$  $(m = 0, \pm 1, \pm 2, \cdots)$ owing to the boundary condition $\psi_{k}(x+Na) = \psi_{k}(x)$. The range for the momentum $k$ is given by $-\pi/a \leq k \leq \pi/a$ since if the Bloch condition holds for $k$ it also holds for $k^\prime = k + 2\pi m/a$. One has from the Schr\"{o}dinger equation that $\psi _{-k}(x) = \psi^{*}_{k}(x)$ and so the negative values of $k$ do not give rise to new solutions. In addition, the values of $m$ are limited to $m = 0, 1, 2, \cdots, N-1$ since the higher values of $m$ do not generate any new solutions and so we have only $N$ solutions per band. In fact, one needs an additional label, a band index, to be added to the Bloch function to describe which band the function belongs.

\subsection{Atom-atom interactions and disordered potentials}

The Bloch form for the condensate wavefunction  $\psi_{\textbf{k}}(\textbf{r})$ given by (35) is appropriate for noninteracting particles in a perfect crystal for $D=3$. However, the condensate (35) does not remove the $1/k^2$--singularity for finite $\textbf{q}_{0}$ for spatial dimensions $D\leq 2$ and $T>0$, which is required for the existence of a BEC for systems of noninteracting or interacting particles when embedded in periodic or in disordered potentials. It is interesting that the removal of the $1/k^2$--singularity leads to localization.  Therefore, we suppose that the inclusion of an interparticle potential, given by the third term in the RHS of (1), gives rise to a BEC provided $\textbf{q}_{0}$ is arbitrarily small and so (35) ceases to be of the Bloch form. It is interesting that this behavior is equivalent to supposing a BEC that is a linear superposition of condensates with differing discrete, translational motion and so the resulting condensate wavefunction is given by
\begin{equation}
\psi(\textbf{r})=  \sum _{\textbf{k}}\hspace{0.0in}^\prime  \alpha_{\textbf{k}} e^{i \textbf{k}\cdot \textbf{r}} u_{\textbf{k}}(\textbf{r}),
\end{equation}
where the prime in the sum indicates that $|\textbf{k}|< |\textbf{q}_{0}|/2$. The existence of a point of accumulation, or limit-point, at $\textbf{k} = 0$ is what is required for the existence of a BEC for systems with spatial dimensionality  $D\leq 2$.  Expression (38) represents a sort of wave packet constructed not by means of plane waves but instead by means of plane waves modulated by a Bloch state, which still leads to an expansion in terms of plane waves as the Fourier series given by (3). It is interesting that such types of states have been used in the study of the dynamics of electron wave packets in crystals \cite{PS09}. In addition, the wave packets so prepared remain in the same band at later times, in our case the lowest energy band, and are referred to as Bloch-type states \cite{PS09} since even though they are not Bloch states, viz. there is no single momentum $\hbar \textbf{k}$ such that $\psi( \textbf{r} + \textbf{t}_{m}) = e^{i \textbf{k}\cdot \textbf{t}_{m}}\psi( \textbf{r})$ even though $u_{\textbf{k}}(\textbf{r} + \textbf{t}_{m}) = u_{\textbf{k}}(\textbf{r})$. The momentum associated with the condensate (38) is
\begin{equation}
\int \textup{d} \textbf{r} \psi^\ast(\textbf{r})(-i \hbar \nabla) \psi(\textbf{r}) = N_{0}\hbar \sum_{\textbf{k}}\hspace{0.0in}^\prime \textbf{k} |\alpha_{\textbf{k}}|^2 = 0,
\end{equation}
where the average momentum of the condensate is zero, that is, we choose the system with respect to which our condensate of $N_{0}$ particles is at rest and so $|\alpha_{\textbf{k}}|^2 = |\alpha_{-\textbf{k}}|^2$. Similarly, the kinetic energy $T$ associated with the condensate (38) is
\begin{equation}
T =  \int \textup{d} \textbf{r} \psi^\ast(\textbf{r})(- \frac{\hbar^2 \nabla^{2}}{2 \mu}) \psi(\textbf{r})=N_{0} \sum_{\textbf{k}}\hspace{0.0in}^\prime \frac{\hbar^2 k^2}{2\mu} |\alpha_{\textbf{k}}|^2 +  N_{0} \frac{\hbar^2 k_{0}^2}{2\mu} \sum_{\textbf{k}} \hspace{0.0in}^\prime  |\alpha_{\textbf{k}}|^2 \sum_{n=-\infty}^{\infty} n^2 |\xi_{\textbf{k} + n \textbf{q}_{0} }|^2,
\end{equation}
where $\mu$ is the particle mass and use has been made of the normalization conditions $\sum_{\textbf{k}} \hspace{0.0in}^\prime  |\alpha_{\textbf{k}}|^2 = 1$ and $\sum_{n=-\infty}^{\infty} |\xi_{\textbf{k} + n \textbf{q}_{0} }|^2 =1$.

One would expect that the kinetic energy $T$ of the condensate should be rather small, which suggest neglecting the second term on the RHS of (40). This corresponds to replacing $u_{\textbf{k}}(\textbf{r})$ in (38) by a constant. Accordingly,
\begin{equation}
\psi(\textbf{r})\approx  \sum _{\textbf{k}}\hspace{0.0in}^\prime  \alpha_{\textbf{k}} e^{i \textbf{k}\cdot \textbf{r}}.
\end{equation}
Note that the BEC (41) is not a Bloch function since $\psi(\textbf{r} + \textbf{t}_{m}) \neq \psi(\textbf{r})$. In addition, for cases where the sum over \textbf{k} is associated with a nonisolated singularity at $\textbf{k} =0$, such a BEC is equally applicable to Anderson localization of ultracold atoms in a disordered potential.

\subsection{BECs in $1D$ crystals}

Consider the BEC in the lowest, or first Brillouin zone, energy band with macroscopic occupation in the first, say, $2 M +1$, lowest energy states, viz. with $k=0, \pm 2\pi/Na, \pm 4\pi/Na, \cdots, \pm 2\pi M/Na$ with $0 < M < N/2$. Thus, the BEC (41) becomes
\begin{equation}
\psi(x)  = \sum_{m= -M}^{M} \alpha_{m} e^{ 2\pi m x i/Na},
\end{equation}
where $\alpha_{m} =\alpha_{-m}$ with normalization condition $\frac{N_{0}}{L} = \sum_{m= -M}^{M} |\alpha_{m}|^2$. Therefore, in the domain $|ka| < 2 \pi M/ N$, our condensate possesses macroscopic occupation of $2M$ single-particle momentum states. Now $\psi(x)$ in (42) satisfies the periodic boundary condition $\psi(x+L) = \psi(x)$; however, $\psi(x)$ does not satisfy the Bloch condition, viz., $\psi(x+ l a)  \neq \psi(x)$ for $0 \leq l <N$.   Notice that in the limit $N \rightarrow \infty$, $M \rightarrow \infty$  such that $ M/N\rightarrow K$,  one has, in any arbitrary neighborhood of the lowest energy state with $k=0$, an unlimited number of nonvanishing $\alpha_{m}$ with $k=0$ an nonisolated singularity. This behavior is what is required to remove the $1/k^2$--singularity that appears in the Bogoliubov inequality in order for the crystalline system to possess a BEC in $1D$.

\subsection{BECs in disordered potentials}

The general expression of BECs in a periodic potential is given by
\begin{equation}
\psi_{\textbf{k}}(\textbf{r}) = \sqrt{\frac{N_{0}}{V(D)}} \sum_{n_{1}, n_{2}, \cdots = - \infty}^{\infty}  \xi_{\textbf{k}+ n_{1} \textbf{q}_{1} + n_{2} \textbf{q}_{2} + \cdots } \hspace{0.1in} e^{i (\textbf{k}+ n_{1} \textbf{q}_{1} + n_{2} \textbf{q}_{2} + \cdots)\cdot\textbf{r}} \equiv  e^{i \textbf{k}\cdot \textbf{r}} u_{\textbf{k}}(\textbf{r}),
\end{equation}
with $u_{\textbf{k}}(\textbf{r}) = u_{\textbf{k}}(\textbf{r} +\textbf{t}_{m})$ for any primitive lattice translation vector $\textbf{t}_{m}$ and where the set of vectors $\{ \textbf{q}_{i}\}$ are in the reciprocal lattice space and so $e^{\textbf{q}_{i}\cdot \textbf{t}_{m}} =1$ for $i=0, 1, 2, \cdots$. The generation of macroscopic occupation in the momenta states in Eq. (43), which occurs, albeit, even in the presence of arbitrarily weak two-body interactions, is a direct consequence of the symmetry breaking term (5) and the supposition that the single-particle momenta states $\textbf{0}, \pm \textbf{q}_{1}, \pm \textbf{q}_{2}, \pm \textbf{q}_{3}, \cdots$ are macroscopically occupied. Of course, if the interparticle potentials are not negligible, then the BEC cannot be of the Bloch form (43). Therefore, in the presence of interparticle interactions, not all the vectors $\{ \textbf{q}_{i}\}$ are in the reciprocal lattice space, in which case the BEC can still be expressed in the form of Eq. (43) except that now $u_{\textbf{k}}(\textbf{r}) \neq u_{\textbf{k}}(\textbf{r} +\textbf{t}_{m})$ and so the BEC is not of the Bloch form. The latter is what one would expect for nonperiodic or disordered potentials for spatial dimensions $D =3$. It is clear that for $D\leq 2$, some of the vectors $\{ \textbf{q}_{i}\}$ must vanish in a limiting process in order to remove the $1/k^2$--singularity in the Bogoliubov inequality (26) required for the existence of a BEC in spatial dimension $D\leq 2$.

It is important to remark that the existence of a BEC for $D\leq 2$, which requires the removal of the $1/k^2$--singularity by means of a point of accumulation at $k=0$, is precisely the same for both noninteracting atoms in a disordered potentials as well as for interacting atoms in the absence of an external potential. Therefore, the mathematical form of the BEC in $D\leq2$ and $T>0$ for noninteracting particles in a disorder potential is indistinguishable from that of interacting atoms in a uniform medium.

\section {Anderson localization of ultracold atoms}

Anderson localization of matter waves has been observed with cold atoms from a noninteracting BEC in a one-dimensional disordered potential generated by a laser speckle pattern \cite{B08} and where the quasi-periodic lattice is the result of the addition of noncommensurate optical periods \cite{RE08}. The experiment consists in releasing a BEC in the $1D$ disordered optical potential, where all the noninteracting atoms are originally in the same single-atom wavefunction, viz. the localized condensate. The atomic wavefunction initially expands and subsequently stops expanding and the resulting wavepacket has wings that decay exponentially or as a power-law \cite{B08} and exponentially or Gaussian-like \cite{RE08}.

The condensate of a nonideal Bose gas in $1D$ and in the absence of an external potential is not given by the macroscopic occupation of a single momentum state, as is the case in an ideal Bose gas in $D=3$, owing to the $1/k^2$--singularity in the Bogoliubov inequality. Actually, for spatial dimensions $D\leq 2$, the condensate must to be nonuniform \cite{MA80} and localized as shown above. Therefore, both noninteracting bosons in disordered potentials and interacting bosons in the absence of external potentials give rise to Anderson localization. It would be interesting if one could experimentally discern the relative contribution to localization by the disordered potential and by the interactions among atoms. In particular, if localization would persist in an ultracold, non-dilute atomic $1D$ gas on expansion, where atom-atom interactions cannot be neglected, even in the absence of a disordered potential, viz., in the absence of any external potential. The experimental proof that a theory, based on the requirements imposed by the Bogoliubov inequality, allows for localization with repulsive atom-atom interactions present would lend some support to the original work of Anderson of a sudden phase transition from conductor to insulator via the degree of disorder in the material \cite{PWA58}. It is interesting that the Mott metal-insulator transition leads to localization without randomness owing to electron-electron interactions, which is somewhat similar to localization in an interacting Bose gas for $D \leq 2$ in the absence of a disordered potential.

Localization has been studied for expanding BECs in weak random potentials \cite{LSP07}.  The localization of a single particle is treated in the Born approximation and the corresponding Lyapunov exponent, characterizing the spatial asymptotic decay of the  BEC density, is determined by values of the correlation length of the disorder $\sigma_{R}$ and the high-momentum cutoff at the inverse healing length $1/\xi_{in}$. For $\xi_{in} > \sigma_{R}$, the BEC wave function is exponentially localized, whereas for $\xi_{in} < \sigma_{R}$, the spatial decay is algebraic \cite{LSP07}. For the speckle potential considered, the Fourier transform of the correlation function vanishes for momenta $k> 2 \sigma_{R}^{-1}$ resulting in a vanishing Lyapunov exponent for $k> \sigma_{R}^{-1}$ in the Born approximation. In one-dimensional elastic scattering, the particle wave vector $k$ in forward scattering remains unchanged or changes sign in backscattering resulting in a momentum transfer of $2k$. Therefore, the study of localization for $k > \sigma_{R}^{-1}$ requires going beyond the Born approximation. The higher order corrections to the Born approximation give rise to "effective mobility edges" at $k=p \sigma_{R}^{-1}$, where $p$ is an integer that characterizes the successive corrections to the Born approximation \cite{EG09, PL09} . It is interesting that these higher-order terms of the Born series are necessary even for $k< \sigma_{R}^{-1}$.

The atom-atom interaction in the Hamiltonian (1) requires macroscopic occupation in momenta $\textbf{k} = n \textbf{k}_{0}$, where $ n= 0, \pm 1, \pm 2, \cdots $, owing to linear momentum conservation, if there is macroscopic occupation in the two momenta states $\textbf{k} = 0, \textbf{k}_{0}$. Therefore, for $1D$, forward scattering of momentum $k$ requires backward scattering into the momentum state $-k$.  Our example of a BEC in $1D$ given by (6) reflects such requirements, which is a signature that one is dealing with a nonperturbative feature of the interparticle potential. Notice that in obtaining the BEC density from the BEC wavefunction (6), one does not suppose that the phases for different momenta are uncorrelated since such supposition would give rise to a constant BEC density, which would be equivalent to supposing macroscopic occupation in a single momentum state and thus to a uniform BEC. The assumption that localized function for a given momentum $k$ are uncorrelated is made in the case of random potentials \cite{LSP07}.

The sum (3) representing the condensate wave function can be approximated arbitrarily well by an integral in any spatial dimension. This is especially important for $D\leq 2$ where the BEC cannot occur in a single-particle momentum state but must be accompanied by a point of accumulation of single--particle momenta condensates at $k=0$ and so the BEC density is spatially nonuniform. For instance, the example for $1D$ given by the sum (6) with its corresponding approximate value given by integral (7). For the case $\nu=1/2$, the integrand of the Fourier cosine transform in (7) possesses two simple poles at $k= \pm i\kappa$ and so for $x > 0$ ($x < 0$), one calculates the integral by closing the contour on the upper (lower) half-plane thus enclosing the simple pole at $k=i \kappa$ ($k= -i \kappa$ ) that yields result (12). Therefore, the long-tail behavior of the BEC density is determined by the singularity in the complex $k$-space closest to the real axis. Note that for $\nu > -1/2$, the nature of the singularities of the integrand in (7) is generally branch points or poles of higher orders at $k= \pm i\kappa$. Nonetheless, the asymptotic formula is still given by an exponential decay, which is a property of the Bessel function $K_{\nu}(z)$ that tends exponentially to zero as $ z \rightarrow \infty$ through positive values. An exponentially decaying BEC density is a direct consequence of Fourier transform $\varphi (k)$  of $\psi(x)$ given by a meromorphic function of $k$, viz., $\varphi (k)$  is analytic except at a set of isolated points, e.g., ratios of rational functions of $k$. If a pole of $\varphi (k)$ is off the imaginary axis, then the asymptotic exponential decay of $\psi(x)$ is modulated by sinusoidal functions, which would represent a remnant periodicity in the system.

In Section II.A, we mentioned examples of BEC in 1$D$ with algebraic localization, viz.,
\begin{equation}
\psi(x) \propto \frac{1}{(x^2 + a^2)^{\nu + 1/2}} = \frac{1}{ (2a)^{\nu }\sqrt{\pi}  \Gamma (\nu + 1/2)} \int_{-\infty}^{\infty}\textup{d}k \hspace{0.03in} |k|^{\nu} K_{\nu}(a|k|) \textup{cos} (kx)
\end{equation}
for $\nu > -1/2$, $a>0$, and $ -\infty < x < \infty$. The function $z^{\nu} K_{\nu} (z)$ is an analytic  function of $z$ for $\nu= n+1/2$,  ($n= 0, 1, 2, \cdots$), but the origin $z=0$ is a branch point singularity for $\nu \neq n+1/2$,  ($n= 0, 1, 2, \cdots$) \cite{W66}. However, the function that appears in the integrand in (44), viz., $|z|^{\nu} K_{\nu}(a|z|)$, is not an analytic function of $z$ since $|z|$ is not an analytic function of $z$. In fact, $K_{\nu}(z)$ is an analytic function of $z$ throughout the $z$-plane cut along the negative real axis since $z=0$ is a branch point singularity.  According to the Riemann--Lebesgue lemma, the Fourier representation of the BEC $\psi(x)$ goes to zero as $x\rightarrow \infty$ and so the large distance behavior of the BEC is determined by the small $k$ behavior of its Fourier transform $\varphi(k)$. Our example (44) of algebraic decay suggests that the behavior of the BEC $\psi(x) \propto 1/(x^2 + a^2)^{\nu + 1/2}$  is determined as $x\rightarrow \infty$ by a branch point singularity at $k=0$. Note also that the value of the BEC $\psi(x)$ for $x \approx 0$ is determined by the behavior of its Fourier transform $\varphi(k)$ over a finite range of values of $k$ near $k=0$ for $\nu > 0$ since $z^{\nu}K_{\nu}(z) \rightarrow 2^{\nu -1} \Gamma(\nu)$ as $z\rightarrow 0$ for $\nu >0$. However, for $\nu=0$, the range of values that contributes to the integral is very small and quite close to $k=0$ owing to $K_{0}(z) \rightarrow -\ln z$ as $z\rightarrow 0$. For instance, the contribution to $\psi(0)$ by the integral (44) for $\nu=1/2$, is 63\% from the region $0 \leq k  < 1/a$ and 37\% from the region $k>1/a$. On the other hand, for $\nu =0$, the contribution to $\psi(0)$ is 79\% from  $0 \leq k  < 1/a$  and 21\% from  $k>1/a$. Note, however, that for the exponentially decaying BEC (7),  $\psi(x) \propto (\kappa |x|)^{\nu} K_{\nu}(\kappa |x|) \rightarrow 2^{\nu -1} \Gamma(\nu)$ as $\kappa|x| \rightarrow 0$ for $\nu >0$, which can become arbitrarily large owing to the simple pole in $\Gamma(\nu)$ at $\nu=0$. However, for $\nu =0$, $\psi(x) \propto K_{0}(\kappa |x|) \rightarrow -\ln (\kappa|x|)$ as $\kappa|x| \rightarrow 0$. This logarithmic divergence at $x=0$ is a direct consequence of the large values of the momentum $k$, which results in the logarithmic divergence of the integral (7). This differs somewhat from the study \cite{LSP07} that suggests that, in general, it is the contribution of waves with very small $k$ that is important for the accurate determination of $\psi(x)$ in the center of the localized BEC. It may be, however, that the momentum $k$ that appears in the Fourier integral (Eq. (6) in the first of Ref. 30) may not be so directly connected with the variable $k$ that appears in the BEC density (Eq. (8) in the first of Ref. (30)) via the stationary, long-time momentum distribution $\mathcal{D}(k)$ and the localized function $\phi_{k}(z)$ of the plane-wave component $e^{ikz}$.

It is interesting that the behavior of the BEC density at large distances is determined by momenta near the high-momentum cutoff $k_{c}$ owing to the localization of the independent $k$ waves \cite{LSP07}. For $\xi_{in} >\sigma_{R}$, the Lyapunov exponent has a finite lower bound that leads to a BEC density that is exponentially localized. On the other hand, for $\xi_{in} <\sigma_{R}$, there is no such finite lower bound and so the localization is algebraic \cite{LSP07}. It is important to remark that the Bogoliubov inequality requires a limit point (or accumulation point) of plane wave momenta at $k=0$ for the existence of a BEC for $D\leq 2$ and that the high-momenta waves do contribute to the limit point and so to the removal of the $1/k^2$--singularity.

\section{The Gross-Pitaevskii Equation}

The numerical results presented in Ref. (30) for the dynamic behavior of the BEC are based on the time-dependent Gross-Pitaevskii equation (GPE) \cite{EPG61}. The GPE represents a mean-field description of the ground state and it is obtained by finding an extremum (a minimum) of the energy as a functional of the BEC wave function \cite{DGP99}. The time-independent GPE for a conserved number of particles corresponding to the Hamiltonian (1) is
\begin{equation}
\frac{-\hbar^2 }{2m} \nabla^2 \psi(\textbf{r}) +   V_{ext}(\textbf{r}) \psi(\textbf{r}) + \psi (\textbf{r})  \int d\textbf{r}^\prime   [V(\textbf{r} -\textbf{r}^\prime) +  V(\textbf{r}^\prime -\textbf{r})] |\psi(\textbf{r}^\prime)|^2 -\mu \psi(\textbf{r}) =0,
\end{equation}
where $\mu$ is the chemical potential.

It is important to remark that the dynamical symmetry-breaking term (5) that requires macroscopic occupation of many single-particle momentum states is determined solely by  the two-body interaction potential $V(\textbf{r} - \textbf{r}^\prime)$ and not at all by the external potential $V_{ext}(\textbf{r})$. Accordingly, the consistency proviso that makes the symmetry-breaking term (5) vanish, viz., that the operator $\hat{\varphi}^\dag(\textbf{r})$ be orthogonal to both $\psi(\textbf{r})$ and $\chi(\textbf{r})$, does not follow from the GPE (45). In fact, for the ground state, our consistency proviso, which requires the last two terms in (45) be orthogonal to the operator $\hat{\varphi}^\dag(\textbf{r})$ , requires, therefore, that the sum $\frac{-\hbar^2 }{2m} \nabla^2 \psi(\textbf{r}) + V_{ext}(\textbf{r}) \psi(\textbf{r})$ in the GPE (45) be also orthogonal to $\hat{\varphi}^\dag(\textbf{r})$. The kinetic energy term is certainly orthogonal to $\hat{\varphi}^\dag(\textbf{r})$ since the kinetic energy is diagonal in the single-particle momentum representation; however, the term $V_{ext}(\textbf{r}) \psi(\textbf{r})$ in (45) is not diagonal in the single-particle momentum representation and, therefore, the Fourier components $V_{ext}(\textbf{k})$ of $V_{ext}(\textbf{r})$ must be in the set $\{\textbf{k}'\}$ of condensate vectors, that is, $\textbf{k} \in \{\textbf{k}'\}$.

For instance, for the BEC in the one-dimensional harmonic trap of Sec. II B,
\begin{equation}
V_{ext}(x) \psi(x) = \sqrt{\frac{N_{0}}{L}} \sum_{q} e^{iqx}  \sum_ {k} V_{ext}(k) \xi_{q-k},
\end{equation}
where $L$ is the length of the one-dimensional ``box." Our consistency proviso requires that (46) be orthogonal to  $\hat{\varphi}^\dag(\textbf{r})$; therefore, not only  $q \in \{k^\prime\}$ but also $k \in \{k^\prime\}$ since otherwise $\xi_{q-k}$ would vanish since $\xi_{q-k}$ is nonzero only for  $(q-p) \in \{k^\prime\}$. Accordingly, the Fourier coefficient $V_{ext}(k)$  of $V_{ext}(x)$ cannot have any nonzero Fourier components outside of the single-particle momenta that constitutes the condensate, viz.,  $\{k'\}$. The Fourier expansion of the harmonic trap is given by
\begin{equation}
x^2 =  \frac{\pi^2}{3k_{0}^2} + 4 \sum_{n=1}^{\infty} (-1)^n  \frac{\textup{cos}(k_{0} n x)}{k_{0}^2 n^2}    \hspace{0.5in}  -\pi/k_{0} \leq x  \leq \pi/k_{0}
\end{equation}
with nonzero coefficients for $k = nk_{0}$, $n=0, \pm 1, \pm 2, \cdots$, which are the same single-particle momenta with macroscopic occupation of the condensate wave function $\psi_{0}(x)$ given by (15).

It should be noted that we are considering the restrictions placed on BECs by the Bogoliubov inequality at finite temperatures. The determination of the condensate wave function requires us to minimize the Helmholtz free energy with respect to the condensate wave function for fixed density and temperature. However, such thermodynamic potential is not available. Therefore, the study of BECs given by the GPE does not suffice since the GPE describes the ground state (zero temperature) of bosonic systems when all the particles are in the condensate. Nonetheless, we have shown that the symmetry breaking term (5), together with the GPE, imposes conditions on the momenta of the Fourier coefficients of the external potential.

\section{Supersolid BEC}

The analysis of the possible existence of BEC for $D\leq 2 $ in the previous sections was based on two-particle, local interactions. It was shown that for periodic potentials, the condensate wavefunction can be of the Bloch form only for $D = 3$. Clearly, local interparticle potentials cannot give rise to a BEC of the Bloch form for $D=2$ since the $1/k^2$--singularity in the Bogoliubov inequality cannot be removed and still preserve the Bloch form for the BEC. 

The $k^2$ behavior of the double commutator (21) follows from the kinetic energy term of the Hamiltonian since local interparticle potentials $\hat{V}$ do not contribute to the Bogoliubov commutator, viz., $\langle[[\hat{C}, \hat{V}], \hat{C}^{\dag}]\rangle = 0$. It is interesting that the latter is not the case for nonlocal potentials \cite{MA71}. If, for instance, the two-particle potential is a sum of a local and a nonlocal potential, then the former potential does not contribute to the Bogoliubov commutator while the latter does and if the decay of the nonlocal potential with distance is sufficiently slow, then $\langle[[\hat{C}, \hat{V}], \hat{C}^{\dag}]\rangle \propto k^{2-\epsilon} $ with $\epsilon > 0$ as $k\rightarrow 0$ \cite{MA71}. Therefore, the symmetry breaking term (5) allows a BEC of the Bloch form for $D=2$ in the presence of an infinitely long-range nonlocal potential between the condensate atoms.

Recently, a Dicke quantum phase transition was realized in an open system formed by a BEC coupled to an optical cavity that gives rise to a self-organized supersolid phase \cite{BG10}. It is interesting that the phase transition is driven by infinitely long-range interactions between the condensed atoms. The analogy of that work to the Dicke model is based on the interaction Hamiltonian that gives rise to a coupling of the pump and cavity fields to the zero-momentum states of the atoms to the symmetric superposition of atomic states that carry an additional unit of photon momentum. This is quite analogous to our dynamically generated symmetry breaking term that allows condensation in atomic states that are integer multiples of a given condensate momentum.

\section{Summary and conclusion}

In random potentials, the underlying mechanism for AL is the suppression of particle transport due to destructive interference. The intriguing question is if such mechanism is undermined by the presence of interparticle interactions. For bosons, we have seen that the generation of periodic BECs is a direct consequence of the dynamical symmetry-breaking term in the Hamiltonian that results from the macroscopic occupation of just two single-particle momentum states, viz., $\textbf{k} = 0, \textbf{k}_{0}$. However, such types of BECs, albeit allowed for $D=3$, violate the Bogoliubov inequality for $D\leq 2$. Therefore, the presence of atom-atom interactions requires that $\textbf{k}_{0} \rightarrow 0$ and thus $\textbf{k} =0$ becomes an accumulation point of condensates. Note that for $D=1$, that suffices to remove the $1/k^2$--singularity in the Bogoliubov inequality. However, for $D=2$, the removal of the $1/k^2$--singularity requires augmenting the set $\{\textbf{k}_{0}\}$ of condensed states so that the removal of the $1/k^2$--singularity occurs for all approaches to the origin $\textbf{k} =0$.

Finally, the existence and nature of BECs for $D\leq 2$ for systems with noninteracting atoms in a disordered potentials or for systems in a ``box" with only atom-atom interactions is precisely the same. Therefore, it may be difficult to discern the individual contributions to localization owing to the strength of the disordered potential or the strength of the interparticle potentials. Experiments with ultracold atomic gases, where the effect of the differing interactions may be easily controlled, will certainly help resolve these very important theoretical questions.

\begin{newpage}
\bibliography{basename of .bib file}

\end{newpage}

\end{document}